\documentclass[runningheads]{llncs}

\usepackage[hidelinks]{hyperref}
\usepackage{url}
\usepackage{graphicx}
\usepackage{booktabs}
\usepackage{makecell}
\usepackage{listings}
\usepackage{color}
\usepackage{amsmath}
\usepackage{multirow}

\newcommand{\ignore}[1]{}

\begin{document}
\title{A Prototype of Serverless Lucene}

\titlerunning{}
\authorrunning{}

\author{Jimmy Lin}

\institute{David R. Cheriton School of Computer Science\\
University of Waterloo}

\maketitle

\begin{abstract}
This paper describes a working prototype that adapts Lucene, the world's most popular and most widely deployed open-source search library, to operate within a serverless environment in the cloud.
Although the serverless search concept is not new, this work represents a substantial improvement over a previous implementation in eliminating most custom code and in enabling interactive search.
While there remain limitations to the design, it nevertheless challenges conventional thinking about search architectures for particular operating points.

\end{abstract}

\section{Introduction}

Individual servers, whether physical or virtual, form the building blocks of search engines.
Persistent services running on a cluster of machines cooperate to provide search functionality, typically in a replicated, document-partitioned architecture~\cite{Barroso03,Baeza-Yates_etal_2007,Dean_WSDM2009}, where each component is responsible for a subset of the entire document collection.

This design has two important implications:\
First, the services must exist as long-running processes, ready to handle requests at any moment.
Second, scaling up and down in response to the query load must be performed at the granularity of {\it servers}, usually through replication and load balancing.
Modern multi-core servers with SSDs can support hundreds of queries per second, and thus a server-based provisioning scheme might not be an ideal match for a particular query load.
For example, given a configuration, performance will steadily degrade as query load increases, until more replicas are brought online, at which point query performance will improve once load is redistributed.
Setting aside the engineering complexities involved in dynamic scaling, this performance profile leads to inconsistent user experiences.
A server-based model also presents a floor on resource consumption, since a server must always be running, even if the load does not require an entire server.
For most ``modest'' search installations, this constitutes significant over-provisioning.
As a concrete example, circa 2016, CiteSeerX received nearly 100,000 queries per day~\cite{balog2016overview}, which translates into only 3.5 queries per second if we assume that all those queries arrive within an interval of eight hours.
Thus, even dedicating two servers (to provide robust failover) represents wasteful resource usage.

The development of cloud technologies can be characterized as continuing disaggregation of computing components.
In the early days, the cloud meant dynamic, readily-available, easy-to-provision virtual machines.
Today, however, there exists a myriad of services, offered by all the major cloud providers, that deliver computing capabilities in a much more fine-grained manner under a pay-as-you-go model.
One particularly interesting development is the rise of so-called Function-as-a-Service (FaaS) offerings:\
The developer provides a block of code with well-known entry and exit points, and the cloud provider handles all other aspects of execution---for example, provisioning resources to execute those functions, scaling up and down to match a particular load, etc., all under a per-invocation cost model.
Combined with storage- and database-as-a-service offerings, it is now possible to write end-to-end serverless applications where the abstraction of a server is completely absent.

Researchers have explored serverless architectures for a variety of applications~\cite{Jonas_etal_SoCC2017,Kim_Lin_WoSC2018,Hellerstein:1812.03651:2018,Fouladi_etal_2019,cloudburst}, but most relevant to this work is serverless search:
Crane and Lin~\cite{Crane_Lin_ICTIR2017} previously demonstrated a working prototype on Amazon Web Services.
In their design, postings lists are stored in the DynamoDB data store and query execution is handled by Lambda (Amazon's FaaS offering).
Their work demonstrated the feasibility of serverless search, but has a number of shortcomings:
End-to-end query latency was around three seconds, which is not fast enough for interactive search.
Furthermore, their prototype required custom code, which presents barriers to broad adoption.

This work addresses both issues.
The contribution here is a prototype of serverless Lucene dubbed ``Anlessini'':
With the aid of minimal adaptor code, Lucene can be packaged to run in Amazon Lambda, reading index structures stored in S3.
Coupled with the use of DynamoDB as a store for the raw documents, it is possible to build an end-to-end serverless search application that uses minimal custom code.
Furthermore, the application is able to achieve query latencies capable of supporting interactive retrieval.

\section{Serverless Lucene}

An important desideratum of this work is to build on the open-source Lucene search library, which has emerged as the {\it de facto} platform for developing real-world search applications, typically via Elasticsearch, Solr, or other components in the broader ecosystem.
Other than a few commercial search engine companies that deploy custom infrastructure (for which the serverless design would not be of interest anyway), Lucene dominates the search landscape, with deployments at organizations ranging from Bloomberg to Twitter to Wikipedia.
Furthermore, the use of Lucene for academic research has been gaining traction~\cite{Azzopardi_etal_SIGIR2017,Azzopardi_etal_SIGIRForum2018,Yang_etal_JDIQ2018}.
Thus, to increase the potential for broader impact and adoption, the focus is to build \textit{serverless Lucene}---leveraging the existing codebase as much as possible---not just ``generic'' serverless search.

\begin{figure}
\includegraphics[width=0.9\linewidth]{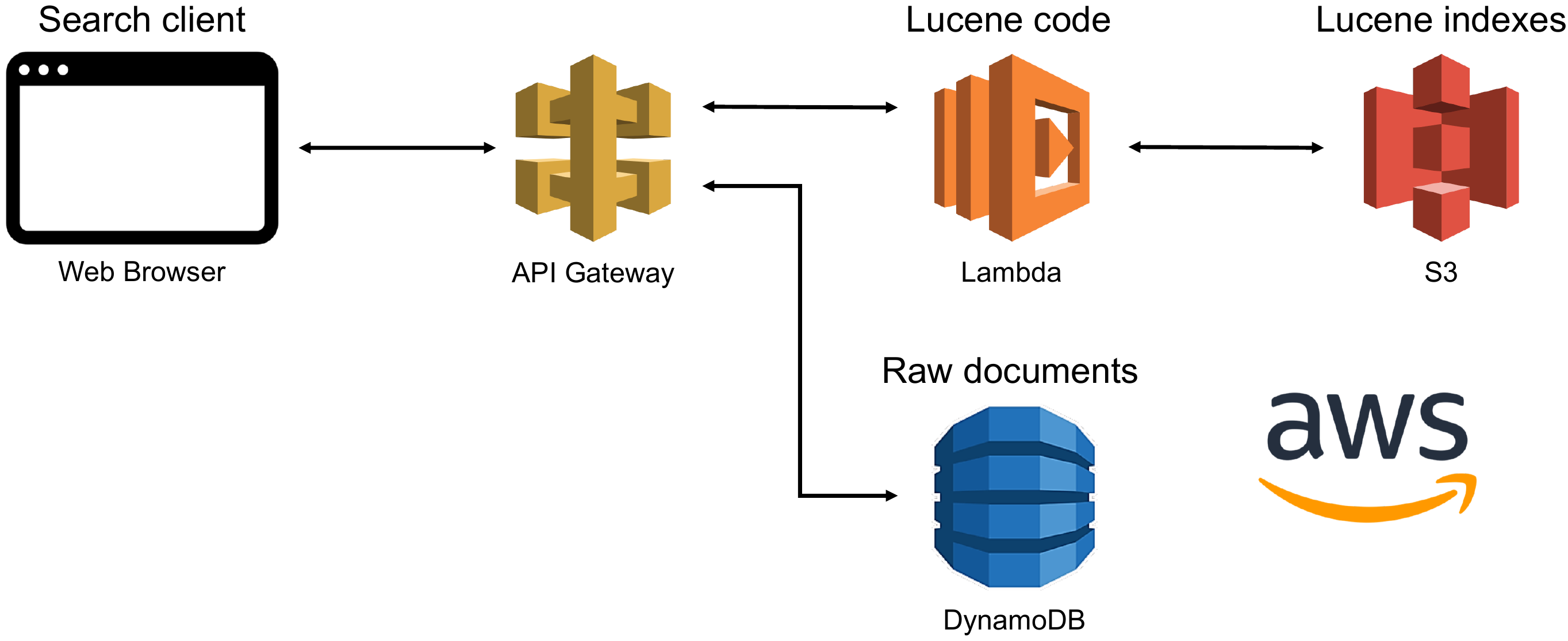}
\caption{Architecture of serverless Lucene ``Anlessini''.}
\label{fig:architecture}
\end{figure}

For this exercise, Amazon Web Services was selected as the cloud platform, although other popular cloud providers offer comparable services.
In general, the design of serverless architectures hinges around the decoupling of state from stateless code.
In the context of search, ``state'' is captured by the inverted index and other related data structures, while query evaluation (i.e., postings traversal to compute rankings) can be considered stateless.
Thus, it is only natural to package Lucene's query evaluation code (\texttt{IndexReader}, \texttt{IndexSearcher}, etc.) into a Lambda function.
The index structures (assumed to have been generated elsewhere) can be held in S3, Amazon's persistent object store.
This yields the basic architecture, dubbed ``Anlessini'', shown in Figure~\ref{fig:architecture}.

How do we ``connect'' Lucene code (running in a Lambda) with index structures stored in S3?
Fortunately, Lucene's \texttt{Directory} interface provides a low-level abstraction for reading index structures (at the level of reading bytes from streams, seeking to different byte offset positions, etc.).
Thus, it suffices to provide a custom \texttt{Directory} implementation built with Amazon's S3 API, and then use this implementation for reading the indexes.
Critically, all other parts of the Lucene query evaluation stack remain unchanged---instead of consuming bytes from a local drive (for example), the bytes are now streamed across the datacenter network from S3.

The final issue is the performance of (remote network) reads from S3.
This is solved by caching; that is, the custom \texttt{S3Directory} implementation reads data into memory and thus the overall design is no different from main-memory search engines, which are quite commonplace today both in the academic literature~\cite{Strohman_Croft_SIGIR2007,Buttcher_Clarke_CIKM2007,Lin_Trotman_IRJ2017} as well as in production deployments~\cite{Dean_WSDM2009,Busch_etal_ICDE2012}.
In order to understand how this caching mechanism interacts with Lambda execution, it is necessary to understand at a high level how Amazon handles FaaS execution.
Behind the scenes, Amazon is provisioning containers to execute the Lambda; it controls how many containers are running to satisfy a particular load, automatically scales up and down the number of containers, and performs load balancing.
Therefore, code execution can either occur on a ``warm'' instance (i.e., already running container) or a ``cold'' instance.
For a ``warm'' instance, query evaluation proceeds without overhead as the index structures have already been loaded into memory; initial execution on a ``cold'' instance, however, carries the additional startup costs associated with populating the cache.
This is not unlike any other in-memory system, and Lambda execution incurs no performance penalty in steady state.

To complete the architecture shown in Figure~\ref{fig:architecture}, there are two more components to describe:\ 
Raw documents are stored in DynamoDB (organized as a simple key--value store) so that they can be accessed as part of the search results.
Finally, all operations are proxied through REST endpoints provided by the API Gateway.
The final product is a full-featured search application accessible to a search client (e.g., in a web browser).

% index size: 692,973,399

The Anlessini serverless Lucene architecture is demonstrated on the MS MARCO passage dataset~\cite{nguyen2016ms} comprising around 8.8M passages.
A simple bag-of-words index, built with the Anserini toolkit~\cite{Yang_etal_JDIQ2018}, occupies around 700 MB storage in S3, and can be comfortably accommodated in a Lambda instance.
Typical queries from the MS MARCO dataset complete in under 300ms (end to end, measured from the browser), which is an order of magnitude improvement over Crane and Lin~\cite{Crane_Lin_ICTIR2017}, and fast enough to support interactive search.
Lambda invocation is charged in terms of memory and time; at the time of writing, each GB/s costs \$0.000016667.
Quantifying cost in terms of round numbers, let's assume a (generous) instance with 2GB memory running for 300ms;
this translates into 100,000 queries per US dollar.
The beauty of the serverless cost model is that query load is entirely fungible---10 QPS for 10,000 seconds or 100 QPS for 1,000 seconds costs exactly the same.

\section{Limitations and Ongoing Work}

This work demonstrates the feasibility of serverless Lucene, an adaptation of the popular open-source search library to run in a serverless environment with minimal code changes.
However, there remain two limitations that represent the focus of ongoing efforts.

First, the architecture assumes static indexes, i.e., there is no mechanism to add or remove documents.
While this is certainly a shortcoming, the current design already fits with existing architectures for periodic batch updates.
Indexes can be built in batch offline, and then bulk loaded into a serving framework~\cite{Leibert_etal_SoCC2011}.
In such a scenario, new indexes can be placed alongside the old, and then the Lambda instances can be refreshed to switch over to the new indexes.
While inadequate for real-time search needs~\cite{Busch_etal_ICDE2012}, this solution can handle (relatively) slowly changing document collections.

Second, the current design assumes that the entire index will fit in memory on a single Lambda instance.
This barrier to scalability, however, can be straightforwardly solved by standard document partitioning practices~\cite{Barroso03,Baeza-Yates_etal_2007,Dean_WSDM2009}, where separate Lambda instances are assigned to different partitions of the document collection.
Given the prototype presented here, building out this design is mostly a matter of software engineering.

While serverless search is not an entirely new concept, serverless Lucene improves upon previous designs and appears to be feasible for practical applications.
Whether such an architecture will gain widespread adoption remains to be seen, but at the very least this design challenges how we think about search architectures for particular operating points.

\section*{Acknowledgments} 

This work was supported in part by the Natural Sciences and Engineering Research Council (NSERC) of Canada.

\bibliographystyle{splncs04}
%\bibliography{serverless}

\end{document}